\documentclass[aps,prl,superscriptaddress,twocolumn,floatfix,a4paper]{revtex4}

\usepackage{graphicx,graphics,epsfig}   
\usepackage{dcolumn}    
\usepackage{bm}         
\usepackage{amsmath}    
\usepackage{verbatim}   
\usepackage{color}      
\usepackage{subfigure}  
\usepackage{times,natbib}
\usepackage{amsmath,amsfonts,amssymb,graphics,graphicx,epsfig,color,times,natbib}

\newcommand{\PPR}[0]{P^{\text{PR}}}

\newcommand{\PL}[0]{P^{\text{L}}}
\newcommand{\Pf}[0]{P^{\text{f}}}

\newcommand{\be}{\begin{equation}}
\newcommand{\ee}{\end{equation}}
\newcommand{\ba}{\begin{eqnarray}}
\newcommand{\ea}{\end{eqnarray}}
\newcommand{\ban}{\begin{eqnarray*}}
\newcommand{\ean}{\end{eqnarray*}}

\newcommand{\ket}[1]{\mbox{$ | #1 \rangle $}}

\begin{document}


\title{Emergence of Quantum Correlations from Non-Locality Swapping}

\author{Paul Skrzypczyk}
 \email{paul.skrzypczyk@bris.ac.uk}
\affiliation{ H.H. Wills Physics Laboratory, University of Bristol, Tyndall Avenue, Bristol, BS8 1TL, United
Kingdom }
\author{Nicolas Brunner}%
\affiliation{ H.H. Wills Physics Laboratory, University of Bristol, Tyndall Avenue, Bristol, BS8 1TL, United
Kingdom }
\author{Sandu Popescu}
\affiliation{ H.H. Wills Physics Laboratory, University of Bristol, Tyndall Avenue, Bristol, BS8 1TL, United
Kingdom } \affiliation{Hewlett-Packard Laboratories, Stoke Gifford, Bristol, BS12 6QZ, United Kingdom}

\date{\today}

\begin{abstract}

By studying generalized non-signalling theories, the hope is to find out what makes quantum mechanics so special. In the present paper, we revisit the paradigmatic model of non-signalling boxes and introduce the concept of a genuine box. This
will allow us to present the first generalized non-signalling model featuring quantum-like dynamics. In
particular, we present the coupler, a device enabling non-locality swapping, the analogue of quantum
entanglement swapping, as well as teleportation. Remarkably, part of the boundary between quantum and post-quantum
correlations emerges in our study.
\end{abstract}


\maketitle

Quantum correlations cannot be ascribed to a local theory \cite{Bell64}, as confirmed by all experiments performed to date \cite{aspect99}. However, Quantum Mechanics (QM) predicts an upper bound on the non-locality of allowed correlations, as shown by Tsirelson
\cite{tsirelson2}. In trying to understand this bound Popescu and Rohrlich \cite{PR} asked whether it was
a direct consequence of relativity -- whether correlations more non-local would lead to signalling -- and
surprisingly found this not to be the case.

This discovery prompted the study of general models, containing more non-locality than QM, but still
respecting the no-signaling principle \cite{barrett}. The ultimate goal of this line of research is to find out
what is special about QM; what distinguishes it from other non-signaling theories. Among the fundamental
questions is the following: What physical principle limits quantum non-locality? This is still unknown today,
but there is no doubt that answering this question will bring deeper understanding of the foundations of QM, as
well as further developments in quantum information science.

Studying the information theoretic properties of generalized non-signaling models has already provided insight
to these questions \cite{BHK,NLcrypto,prsinglet,john}. On the one hand, many astonishing features
of QM, such as no-cloning, no broadcasting and monogamy of correlations, have been shown to be general properties of any non-signaling model \cite{NS,john,barnum}. Hence these properties do not indicate any separation between QM and
post-quantum theories. On the other hand, van Dam \cite{vanDam} and Brassard \textit{et al.}\cite{brassard} showed that particular classes of
post-quantum models allow for a dramatic increase of communication power compared to QM.
Moreover, Linden \textit{et al.}\cite{noah} showed that the same post-quantum theories allow for non-local computation while QM
does not, here providing a tight separation between QM and post-quantum models.

More importantly however, there is one crucial aspect of QM that generalized models have failed to reproduce
until now, namely its dynamics; in particular, the ability to perform joint measurements on two systems, which
is the key ingredient for fascinating quantum processes such as teleportation \cite{TeleportationBennett} and
entanglement swapping \cite{EntSwapping}. In fact, Short \textit{et al.} \cite{short} and Barrett \cite{john} showed that there are no joint
measurements in theories constrained only by no-signaling, thus suggesting the existence of
another fundamental principle inherent to QM, that generalized models fail to capture.

Here we take a new conceptual perspective on generalized non-signalling models, which allows us to implement
joint measurements. We revisit the paradigmatic model of Popescu-Rohrlich (PR) boxes \cite{PR} and introduce the concept of a \emph{genuine box}. This allows us to present a model featuring rich dynamics, such as non-locality swapping, the analogue of quantum entanglement swapping, and
teleportation. Joint measurements are implemented using an imaginary device called a \emph{coupler}. Finally, and probably most surprisingly, we show that the set of quantum correlations partially emerges in our model.

\begin{figure}[b]
  \includegraphics[width=0.8\columnwidth]{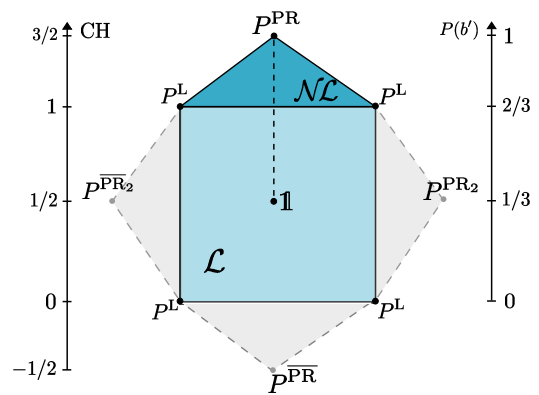}\\
  \caption{The set of allowed states is restricted to (i) the local polytope $\mathcal{L}$ and (ii) the PR box. The other PR boxes are discarded since they are not genuine, and should therefore not be considered for non-locality swapping. The left axis is the CH value. Local states satisfy $0 \leq \vec{\text{CH}}\cdot\vec{P^L} \leq 1$, the CH Bell inequality. The coupler (right axis) is a re-scaling of the CH value (see text). Note that the polytope is 8-dimensional; the figure is a 2-dimensional illustration.}\label{polytope}
\end{figure}

\begin{figure*}[t]
  \includegraphics[width=1.9\columnwidth]{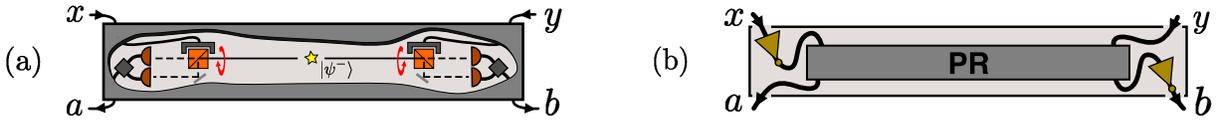}\\
  \caption{Genuine boxes. (a) A `quantum' black-box contains a quantum state and measurement devices (polarizers, detectors). The orientation of the polarizers depends on the input values $x,y$. The genuine part of the box is the quantum state; before performing a joint measurement, one must remove the measurement devices. (b) A non-genuine PR box contains the genuine PR box and classical circuitry. Importantly, upon applying the coupler, one should first open the box, remove the circuitry, and connect the coupler directly to the genuine PR box.}\label{genuine}
\end{figure*}

\emph{Genuine boxes.} As we shall work with generalized non-signaling theories, the quantum formalism is no longer relevant; here
bipartite states are not given by vectors in a Hilbert space but by bipartite joint probability distributions;
i.e. probabilities of a pair of results (outputs) given a pair of measurements (inputs). In other words, quantum
correlations will be replaced by more general ``boxes" ( i.e. input-output devices).

Here we shall focus on the simplest possible scenario, namely the case of two possible measurements for each
party (inputs $x,y \in \{0,1\}$); each measurement providing a binary result (outputs $a,b\in \{0,1\}$). It is
very insightful to think of such a scenario in geometrical terms \cite{pito}. In this approach, a box is viewed as a vector in a space of boxes; the vector's components are the joint probabilities characterizing the box.

The set of boxes that can be obtained from local means only forms a polytope. This local polytope is itself
embedded in a larger polytope, the non-signalling polytope, which contains all boxes compatible with the
non-signalling principle \cite{barrett}. It has 8 non-local vertices, which are all symmetries of the PR
box. The set of boxes attainable by QM also form a convex body, though not a polytope. The
quantum set is strictly larger than the local polytope -- quantum correlations can be non-local -- but strictly
smaller than the non-signalling polytope -- quantum non-locality is bounded by Tsirelson's bound.

Here we will voluntarily restrict the set of allowed boxes, by discarding certain non-signaling boxes. More precisely, we consider the entire local polytope, given by its 16 vertices, the deterministic boxes
\begin{equation}\label{local}
\PL_{\alpha \beta \gamma \delta}(ab|xy) =
\begin{cases}
1 & \text{if} \,\, \text{$a = \alpha x \oplus \beta$} \,\, , \,\, \text{$b = \gamma y \oplus \delta $} \\
0 & \text{otherwise}
\end{cases}
\end{equation} parameterized by $\alpha,\beta,\gamma,\delta \in \{0,1\}$. To this we add a single non-local vertex, the PR box:

\begin{equation}\label{PR}
\PPR (ab|xy) =
\begin{cases}
\frac{1}{2} & \text{if} \,\, \text{$a \oplus b = xy$} \\
0 & \text{otherwise}
\end{cases}
\end{equation} where $\oplus$ is addition modulo 2. The resulting set of boxes forms a polytope (see Fig. 1). The non-locality of a given box is characterized by the Clauser-Horne (CH) value \cite{ch}. Since boxes are considered as vectors, it is convenient to denote the CH value of a box as a scalar product $ \vec{\text{CH}}\cdot\vec{P}(ab|xy) = P(11|00) + P(00|10)+ P(00|01) - P(00|11) $.

The fact that we consider only a single PR box turns out to be a crucial aspect of our model. However, one may
wonder why the other seven symmetries of the PR box are not taken into account. Below we argue that these restrictions have in fact a deep significance, and force us to re-examine the conceptual foundations of this entire line of research. Actually, revisiting the model will turn
out to be highly rewarding, since it will provide it with rich dynamics, a feature proven to be impossible when all non-signalling states are considered on an equal footing \cite{short,john,tonijohn}.

Let us first re-examine the standard ``black box" approach to quantum correlations, where they are stripped back
to their purest form; measurement choices and outcomes are both reduced to single bits of information
\cite{DevIndep}. It is instructive to think about how such a setup would in reality be produced.

The black box consists of a quantum system and measuring devices (see Fig.
2a). For the case of two polarized photons, we would require one polarizer on each
side of the box, with two possible orientations, and detectors recording the
measurement outcome and outputting the corresponding bit. Here, the quantum state is the genuine part
of the box, i.e. the non-local resource; the measurement is then a processing. Indeed, by changing the orientation of the polarizers one can produce many different black-boxes starting from the same initial quantum state, but
they are clearly not genuinely different. Moreover, it is also possible to produce the same black-box by using two different quantum states, subjected to appropriate measurements.

In the case of the PR box there are clearly no quantum states
and polarizers in the box, and so it is more delicate to separate what is genuine in the box from what is not.
Note that as long as we do not need to look inside the box, as is the case in most of the scenarios considered
so far, we need make no distinction between genuine and non-genuine.

However, when dynamics are introduced in the model, things change. Let us first think of how a joint measurement would be implemented in the quantum case; importantly it is performed on quantum particles, and not on measurement outcomes. Thus, one should first \emph{open the box}, remove the polarizers, and then perform the joint measurement on the two particles; i.e. it must be performed on the genuine part of the box, that is on the quantum state and not on the box itself. Clearly, now it is important what the actual quantum state is: two boxes that appeared to be the same while containing
different states and polarizers, may now behave differently.

We argue that for PR boxes, the situation is fully analogous. All 8 PR boxes are not genuinely different, since they can be generated from the single PR box \eqref{PR} by adding classical circuitry (see Fig. 2b). In order to perform a joint measurement, it is crucial to apply the coupler to the genuine part of the box, i.e. to the genuine PR box; all circuitry must be removed.

In the light of the results of Ref \cite{short,john} and of our own results, we argue that the concept of
genuine boxes is not a particularity of the quantum case, but is in fact general to non-signaling models. When all 8 PR boxes are put on
an equal footing, i.e. as all being genuine, then interesting dynamics are forbidden \cite{short,john}. Here we
choose the simplest and most natural possibility: we consider as genuine all local deterministic states
\eqref{local} and we add one single PR box \eqref{PR}. This choice will allow for rich dynamics, as we show
below. Importantly, our model does not restrict the set of valid probability distributions (circuitry is allowed); we only restrict the set of genuine boxes, on which the coupler must be consistently defined.

\begin{figure}[t]
  \includegraphics[width=0.8\columnwidth]{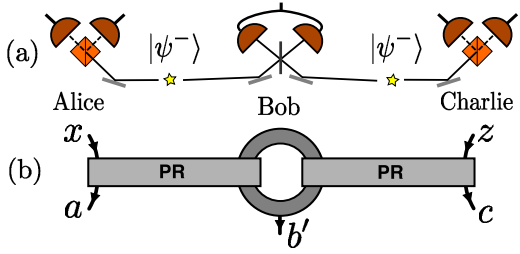}\\
  \caption{(a) Entanglement swapping and (b) Non-locality swapping. The coupler is the analogue of the quantum joint measurement.}
\end{figure}

\emph{Coupler for non-locality swapping.} Let us start by briefly reviewing the quantum protocol of entanglement swapping \cite{EntSwapping}. An observer, Bob, shares a maximally entangled state with both Alice and Charlie, two distant observers. Initially, the global state of the system is
$\ket{\psi^-}_{AB_1}\ket{\psi^-}_{B_2C}$, where $\ket{\psi^-}$ is the
singlet state. The core of the protocol is the ability of Bob to perform a joint measurement of his two particles
($B_1,B_2$). The simplest form of such a measurement is the projection onto the symmetric and antisymmetric subspaces of two qubits, i.e. $M = \left\{|\psi^-\rangle\langle\psi^-|, \openone - |\psi^-\rangle\langle\psi^-|\right\}$. When the protocol is successful (Bob's particles are projected onto $\ket{\psi^-}$), the global state undergoes the transformation $\ket{\psi^-}_{AB_1}\ket{\psi^-}_{B_2C} \rightarrow \ket{\psi^-}_{B_1B_2} \ket{\psi^-}_{AC}$; entanglement has been swapped between $AB_1$ and $B_2C$ to $B_1B_2$ and $AC$. Importantly, it is not until the announcement of Bob's successful joint measurement that Alice and Charlie learn which state they hold and
whether they share any non-local correlations.

It is straightforward to extend the scenario of entanglement swapping to our generalized model. Instead of
entangled quantum states, Bob shares now a non-local box with both Alice and Charlie. To implement non-locality swapping one must first define the coupler, the analogue of a
quantum joint measurement (see Fig. 3). When applied on Bob's two boxes, the coupler encompasses the inputs and outputs and returns a single bit $b'$, i.e. implementing the linear transformation $  P(ab_1|xy_1)P(b_2c|y_2z) \xrightarrow{\chi} P(ab^\prime c|xz) $ \cite{short}.

Now let us define the action of the coupler on two PR boxes:
\begin{multline}
     \PPR(ab_1|xy_1)\PPR(b_2c|y_2z) \xrightarrow{\chi} P(ab^\prime c|xz) \\
    = \begin{cases}
    \ q \PPR(ac|xz) & \text{if} \,\,\text{$b^\prime=0$} \\
    \ (1-q) \Pf(ac|xz) & \text{if} \, \, \text{$b^\prime=1$}
    \end{cases}
\end{multline} With probability $q=P(b^\prime=0)$ Bob succeeds in swapping a PR box to Alice and Charlie. With probability $1-q=P(b^\prime=1)$ the protocol fails, and Alice and Charlie share the failure box $\Pf$. Importantly, relativity imposes that Bob could not signal by applying the coupler. Therefore the (reduced) box of Alice and Charlie must be independent of whether Bob applied the coupler or not, i.e.

\ba
   P(ac|xz)=\sum_{b_1,b_2} P(ab_1|xy_1)P(b_2c|y_2z) = \sum_{b^\prime}P(ab^\prime c|xz) \,.
\ea In case Bob shares a PR box with both Alice and Charlie, one has that $P(ac|xz)=\openone(ac|xz) = \tfrac{1}{4} \: \forall \:a,c,x,z$, the fully mixed state. The requirement that $\Pf$ is an allowed box imposes a limit on the probability of success; here we make the optimal choice $q=\frac{1}{3}$. Thus we have $\Pf(ac|xz) = \frac{3}{2}\left(\openone(ac|xz) - \frac{1}{3} \PPR(ac|xz)\right)$ and $\vec{\text{CH}}\cdot\vec{\Pf}=0$.

Next it must be checked that the coupler acts consistently when applied directly to \textit{any} allowed box; not only when it is applied between two boxes. For example, the output probabilities must be positive when the coupler is connected to both ends of a single PR box. Here we just sketch the argument; the full proof can be be found in Appendix A. The proof is based on the following observation: if Bob, after applying the coupler, learns from Alice and Charlie their respective inputs and outputs,
he should get the same result as if he learned Alice's and Charlie's inputs and outputs first and then applied the coupler. We find that the coupler outputs $b^\prime=0 $ with a probability proportional to the CH value of the box it is applied to, i.e.
\begin{equation}\label{coupler}
    P(b^\prime=0|P(ab|xy)) = \tfrac{2}{3}\vec{\text{CH}}\cdot\vec{P}(ab|xy)\,.
\end{equation}
The constant of proportionality is here crucial, since it ensures that the coupler outputs with a valid
probability when applied to any allowed box (see Fig. 1). Notably, upon applying the coupler directly to a PR box, one always obtains the outcome $b^\prime=0$. This is exactly what happens in the quantum case: when Bob holds a
singlet and performs a joint measurement, he always projects onto $\ket{\psi^-}$.

Note that the inconsistency of the potential coupler presented in Ref. \cite{short} becomes now clear, since it output with a probability equal to the CH value; thus, when applied onto the PR box,
it output with a non-valid probability of $\frac{3}{2}$. Along the same line, it is also clear why our coupler runs
into inconsistencies if we try to reintroduce disallowed (non-genuine) PR boxes; for instance the
anti-PR box (given by $a\oplus b\oplus 1 = xy$) would lead to negative probabilities.

Finally to be consistent, the coupler must take any two genuine boxes to a genuine box. It is straightforward to check that this is the case, by applying the coupler to all pairs of vertices.

\begin{figure}
  \includegraphics[width=0.9\columnwidth]{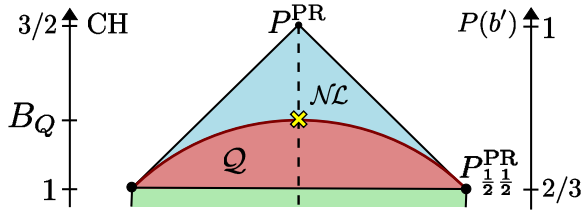}\\
  \caption{In a whole section of the polytope, the boxes useless for non-locality swapping correspond exactly to the set of quantum boxes $\mathcal{Q}$ (red region), characterized by the TLM criteria; for isotropic boxes (dashed line), this corresponds to Tsirelson's bound (cross) $B_Q=\frac{1}{\sqrt{2}}+\frac{1}{2}$, the quantum bound of the CH inequality.}\label{polytope}
\end{figure}

\emph{Emergence of quantum correlations.} The coupler enables \emph{perfect} swapping of two PR boxes; this means that the final state of Alice and Charlie, which is a PR box, is as non-local as the initial states shared by Alice-Bob and Bob-Charlie. Now a natural question to ask is whether Bob, by applying the coupler, can also swap non-locality starting from imperfect boxes. Here we consider a natural section of the polytope, which includes PR, $\text{PR}_2$ (another PR box given by $a \oplus b = xy \oplus x)$), and the identity $\openone$. Thus we have noisy boxes of the form

\ba\label{noisy} \PPR_{\xi,\gamma} \equiv \xi \PPR + \gamma P^{\text{PR}_2} + (1- \xi - \gamma )  \openone \ea
with $\xi + \gamma \leq 1$. Note that these boxes are genuine as long as $\gamma\leq \frac{1}{2}$ \footnote{Note that here we used the (non-genuine) box $\text{PR}_2$ only for mathematical convenience. We could have used a decomposition involving only genuine boxes.}. Of particular interest are isotropic boxes $\PPR_{\xi,0}$, that lie on the line joining the PR box and the identity. One finds that $\vec{\textrm{CH}} \cdot \vec{P}^{\text{PR}}_{\xi,\gamma} = \xi+ \frac{1}{2}$.

Using the linearity of the coupler one can check that when Bob succeeds in swapping non-locality (i.e. he gets $b'=0$) starting from two $\PPR_{\xi,\gamma}$ boxes, the final state of Alice
and Charlie has CH value $\vec{\textrm{CH}} \cdot \vec{P}(ac|xz) = \xi^2 + \gamma^2
+\frac{1}{2}$. Thus, the coupler enables perfect swapping only for noiseless PR boxes; two noisy boxes can only be swapped to an even noisier box.

Remarkably, non-locality can be swapped using two boxes $\PPR_{\xi,\gamma}$ if and only
if $\PPR_{\xi,\gamma}$ is post-quantum; that is iff $\vec{P}^{\text{PR}}_{\xi,\gamma}$ violates the Tsirelson-Landau-Masanes (TLM) inequality
\cite{tsirelson,masanes,landau}, a necessary and (here) sufficient condition for a box
to be quantum. Thus, when the two initial boxes $\PPR_{\xi,\gamma}$ are noisy enough to have been
produced quantum mechanically, the resulting box shared by Alice and Charlie is so noisy as to become local. For isotropic boxes, this condition reduces to $\vec{\textrm{CH}} \cdot \vec{P}^{\text{PR}}_{\xi}> B_Q = \frac{1}{2}+\frac{1}{\sqrt{2}}$, where $B_Q$ is the Tsirelson bound of
the CH inequality.

\emph{Proof.} Boxes $\PPR_{\xi,\gamma}$ useless for non-locality swapping, i.e. leading to
$\vec{\textrm{CH}} \cdot \vec{P}(ac|xz)\leq 1$, are characterized by the relation \ba\label{boundary}  \xi^2 +
\gamma^2 \leq \frac{1}{2}\,\,. \ea  The TLM criteria is written here in the form of Laudau \cite{landau}
\ba\label{landau}\nonumber |E_{00}E_{01}-E_{10}E_{11}|  &\leq&  \sqrt{(1-E_{00}^2)(1-E_{01}^2)} \\&+&
\sqrt{(1-E_{10}^2)(1-E_{11}^2)} \ea where $E_{xy}=P(a=b|xy)-P(a\neq b|xy)$ is the correlator associated to the
pair of measurements $x,y$. For noisy states $\PPR_{\xi,\gamma}$, the four correlators are given by
$E_{00}=E_{01}=\xi+\gamma$ and $E_{10}=-E_{11}=\xi-\gamma$. Inserting these last expressions in \eqref{landau},
we get exactly the relation \eqref{boundary}, which completes the proof.

Let us point out however that not the entire quantum versus post-quantum boundary emerges in this way: on other
sections of the polytope the coupler ceases to swap non-locality before reaching the quantum bound.

\emph{Conclusion and Perspectives.} In summary, we revisited the post-quantum model of PR boxes,
introducing the concept of genuine boxes. This allowed us to to consider a restricted space of non-signalling boxes; this space features much richer dynamics than the full non-signaling space. We presented the coupler, a device enabling non-locality swapping. The coupler also implements teleportation (see Appendix B). Even more surprisingly, quantum correlations partially emerged from the coupler. Though we do not understand its full significance at this stage, we believe this intimate connection is tantalizing, since it links a dynamical process in a natural non-signalling model directly to QM. In the future we plan to investigate further on this link, and look for a fundamental principle
potentially underlying it. Studying other information theoretic tasks from the new perspective
of genuine boxes may help us understand what is so special about QM.

\emph{Acknowledgements.} The authors are grateful to J. Barrett, N. Gisin, V. Scarani and A. J. Short for insightful discussions. S.P. and P.S.
acknowledge support through the UK EPSRC project 'QIP IRC'. N.B. acknowledges financial support by the Swiss
National Science Foundation.

\bibliographystyle{prsty}
\bibliography{C:/BIB/thesis}

\subsection{Appendix A: Deriving the action of the coupler on allowed states}

Here we derive the action of the coupler on any allowed box (see Fig. 5). Since all consistent couplers are linear
functions of the inputs and outputs of a box \cite{short}, it is sufficient here to consider only extremal
boxes.

\begin{figure}[b]
  \includegraphics[width=0.22\columnwidth]{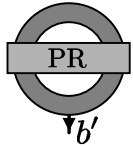}\\
  \caption{Bob applies the coupler directly to a PR box.}
\end{figure}

Let us start with the deterministic ones. Note first that the output of the coupler must be consistent
regardless of the timings of Alice's and Charlie's inputs and of Bob's application of the coupler. That is, if
Bob, after applying the coupler, learns from Alice and Charlie their respective inputs and outputs, he should
get the same result as if he learnt Alice's and Charlie's inputs and outputs and then applied the coupler.
Conditioning the final probability distribution $P(ab'c|xz)$, on Bob learning Alice's and Charlie's inputs and
outputs, one has $P(b^\prime=0|acxz) = \frac{2}{3}$, when $xz\oplus a \oplus c = 0$, and $P(b^\prime=0|acxz)=0$
otherwise. Finally we note that when Bob learns the four values $a,c,x,z$ he knows that he is holding the
extremal local box $\PL_{xazc}$, thus the action of the coupler on the deterministic states is given by
\begin{equation}\label{e1}
    P(b^\prime=0|\PL_{\alpha\beta\gamma\delta}) =
    \begin{cases}
    \frac{2}{3} & \text{if} \,\, \text{$\alpha\gamma\oplus \beta \oplus \delta = 0$} \\
    0 & \text{otherwise}\,.
    \end{cases}
\end{equation}
This again can be understood geometrically -- if the local box is on the facet $\text{CH} = 1$ the coupler
outputs $b^\prime=0$ with probability $\frac{2}{3}$, whilst if the box is on the facet $\text{CH} = 0$ then it
deterministically outputs $b^\prime = 1$.

Next, let us find the action of the coupler on the PR box. In order to do this, we decompose a given probability
distribution in two different ways. We consider the point $P^c(b_1b_2|y_1y_2)$ in the centre of the $\text{CH} =
1$ facet, half way between the PR box and the identity, which can be written as a convex combination of the
extremal boxes in the following ways:
\begin{align}\label{d1}
    P^c(b_1b_2|y_1y_2) &= \frac{1}{8} \sum_{\alpha\beta\gamma} \PL_{\alpha\beta\gamma(\alpha\gamma\oplus\beta)}(b_1b_2|y_1y_2) \\
                        &= \frac{1}{2} \left( \PPR (b_1b_2|y_1y_2) + \openone(b_1b_2|y_1y_2)\right)
\end{align}
the first decomposition \eqref{d1} being the equal sum of the 8 extremal vertices of the upper facet. By
applying the coupler to both decompositions, using the linear action of the coupler on the convex sum and
demanding they give the same probability of outputting $b^\prime$, it is found that
\begin{equation} \label{e2}
    P\left(b^\prime=0| \PPR(b_1b_2|y_1y_2)\right) = 1 \,.
\end{equation}
Thus upon applying the coupler directly to a PR box, Bob always obtains the outcome $b^\prime=0$; exactly as
in the quantum case.

From inspection of equations (\ref{e1}) and (\ref{e2}) it is clear that the coupler outputs $b^\prime=0 $ with a
probability that is proportional to the CH value of the box it is applied to, i.e.
\begin{equation}
    P(b^\prime=0|P(ab|xy)) = \tfrac{2}{3}\vec{\text{CH}}\cdot\vec{P}(ab|xy) \,.
\end{equation}
This constant of proportionality ensures that the coupler outputs a valid probability when applied to any
allowed box. Note that the inconsistency of the potential coupler presented in Ref. \cite{short} becomes now
clear, since it outputted with a probability equal to the CH value; therefore, when applied onto the PR box,
it gave a non-valid probability of $\frac{3}{2}$. Along the same line, it is also clear why the coupler runs
into inconsistencies if we try to reintroduce the seven disallowed non-genuine PR boxes; for instance the
anti-PR, defined by the relation $a\oplus b\oplus 1 = xy$ would lead to negative probabilities.

\subsection{Appendix B: Teleportation.}

When Alice-Bob share a PR box, and Bob holds a
deterministic box $\PL_{\alpha\beta}$, the coupler implements the transformation \ba
\PPR(ab_1|xy_1) \PL_{\alpha\beta}(b_2|y_2) \xrightarrow[b'=0]{\chi}
\PL_{\alpha\beta}(a|x) \,. \ea Therefore the final box held by Alice (given that the joint
measurement succeeded) is $\PL_{\alpha\beta}(a|x)$ (see Fig. 6). Thus, Bob can teleport to Alice any single-party box $P_B(b|y) =
\sum_{\alpha,\beta}p_{\alpha\beta} \PL_{\alpha\beta}(b|y)$, with $\sum_{\alpha,\beta} p_{\alpha\beta}=1$,
which can be seen by using the linearity of the coupler. Here the PR provides the teleportation
channel, as does the maximally entangled state in the quantum protocol.

\begin{figure}[b]
  \includegraphics[width=0.62\columnwidth]{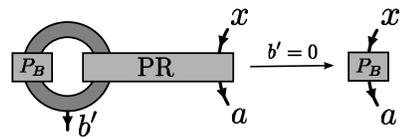}\\
  \caption{Teleportation with the coupler.}
\end{figure}

\end{document}